
\headline={\ifnum\pageno=1\firstheadline\else
\ifodd\pageno\rightheadline \else\leftheadline\fi\fi}
\def\firstheadline{\hfil}
\def\rightheadline{\hfil}
\def\leftheadline{\hfil}

\font\twelvebf=cmbx10 scaled\magstep 1
\font\twelverm=cmr10 scaled\magstep 1
\font\twelveit=cmti10 scaled\magstep 1

\font\tenbf=cmbx10
\font\tenrm=cmr10
\font\tenit=cmti10

\font\ninerm=cmr9

\parindent=1.5pc
\hsize=6.0truein
\vsize=8.5truein

\def\ag{{\cal A}/{\cal G}}
\def\agb{\overline{\ag}}
\def\agr{\ag}
\def\V{V_{\rm kin}}
\def\H{H_{\rm kin}}
\def\Ab{\bar{A}}
\def\tit{\twelveit}
\def\tbf{\twelvebf}

\centerline{\tenbf RECENT MATHEMATICAL DEVELOPMENTS IN}
\baselineskip=16pt
\centerline{\tenbf QUANTUM GENERAL RELATIVITY}
\vglue 0.8cm
\centerline{\tenrm ABHAY ASHTEKAR}
\baselineskip=13pt
\centerline{\tenit Center for Gravitational Physics and Geometry}
\baselineskip=12pt
\centerline{\tenit Physics Department, Penn State, University Park, PA 16802}
\vglue 0.3cm
\centerline{To Appear in the Proceedings of the Seventh Marcel Grossman
Conference}
\vglue 0.8cm \centerline{\tenrm ABSTRACT}
\vglue 0.3cm
\vglue 0.6cm
{\rightskip=3pc \leftskip=3pc \tenrm\baselineskip=12pt\noindent After
a brief chronological sketch of developments in non-perturbative
canonical quantum gravity, some of the recent mathematical results are
reviewed. These include: i) an explicit construction of the {\it
quantum} counterpart of Wheeler's superspace; ii) a rigorous procedure
leading to the general solution of the diffeomorphism constraint in
quantum geometrodynamics as well as connection dynamics; and, iii) a
scheme to incorporate the reality conditions in quantum connection
dynamics.  Furthermore, there is a new language to formulate the
central questions and techniques to answer them.  These developments
put the program on a sounder footing and, in particular, address
certain concerns and reservations about consistency of the overall
scheme.

\vfil
\twelverm
\baselineskip=14pt
\leftline{\twelvebf 1. Introduction}
\vglue 0.4cm

It is well known that quantum general relativity is perturbatively
non-re\-norma\-lizable. Particle theorists often take this to be a
sufficient reason to abandon general relativity and seek an
alternative which has a better ultraviolet behavior in perturbation
theory. However, one is by no means forced to this route. For, there
do exist a number of field theories which are perturbatively
non-renormalizable but are {\it exactly soluble}. An outstanding
example is the Gross-Neveau model in 3 dimensions, $(GN)_3$, which was
recently shown${}^{1}$ to be exactly soluble rigorously. Furthermore,
the model does not exhibit any mathematical pathologies. For example,
it was at first conjectured that the Wightman functions of a
non-renormalizable theory would have a worse mathematical behavior.
The solution to $(GN)_3$ showed that this is not the case; as in
familiar renormalizable theories, they are tempered distributions.
Thus, one can argue that, from a structural viewpoint, perturbative
renormalizability is a luxury even in Minkowskian quantum field
theories. It does serve as a powerful guiding principle for selecting
physically interesting theories since it ensures that the predictions
of the theory at a certain length scale are independent of the
potential complications at much smaller scales. But it is {\it not} a
consistency check on the mathematical viability of a
theory. Furthermore, in quantum gravity, one is interested precisely
in the physics of the Planck scale. The short-distance complications
are now the issues of primary interest.  Therefore, it seems
inappropriate to elevate perturbative renormalizability to a viability
criterion.

Even if one accepts this premise, however, one is led to ask: Are
there reasons to expect that quantum general relativity may exist at a
non-perturbative level? The answer, I believe, is in the affirmative.
There are growing indications from a number of different directions
--computer simulations, canonical quantization and string theory--
that the quantum geometry of space-time would be quite different from
classical geometry${}^{2-4}$. Indeed, there are concrete calculations
that hint at a discrete structure at the Planck scale. The
perturbative treatments, on the other hand, assume the validity of a
continuum picture at all scales. The ultraviolet problems one
encounters may simply be a consequence of the fact that the true
microscopic structure of space-time is captured so poorly in these
treatments. Put differently, if the continuum picture is replaced by a
more faithful one, the ``effective dimension'' of space-time could be
smaller than four, whence the theory could have a much better behavior
non-perturbatively.%
\footnote{${}^\dagger$}{\ninerm \baselineskip=7pt\noindent
If one considers general relativity with two Killing fields, one
encounters a rather dramatic example of a non-perturbative
effect${}^{5}$: while the perturbative Hamiltonian is, as one would
expect, unbounded above, the exact Hamiltonian is a non-polynomial
function of the perturbative Hamiltonian and is in fact {\it bounded
above}. One would therefore expect that this field theory would be
free of the usual ultraviolet difficulties.}

In this article, I will accept this premise and consider a
non-perturbative quantization of general relativity.  This program is
based on canonical quantization. More precisely, one casts general
relativity in a Hamiltonian form, realizes that it is a dynamically
constrained system and uses (an extension of) Dirac's method for
quantization of such systems. It {\it is} true that such canonical
methods require a 3+1 splitting of space-time and therefore lack
manifest covariance. Nonetheless, when applied to diffeomorphism
invariant theories such as general relativity, they also have a number
of advantages. First, they do not require the use of a background
metric or a connection or indeed any background field. In this sense,
they respect the diffeomorphism invariance of the theory in an
essential way. Second, it seems extremely hard to give a mathematical
meaning to other methods such as path integrals in general relativity.
The difficulties are not merely the ``technical'' ones associated with
functional analysis. They are also conceptual, having to do with the
celebrated ``issue of time'' in quantum gravity.  In particular, while
one can extend the key axioms of Osterwalder and Schrader for
Euclidean quantum field theory to diffeomorphism invariant
theories${}^6$, because of the absence of a sensible Wick rotation
procedure, the transition from Schwinger functions to Wightman ones
seems impossible to achieve. Thus, in the Euclidean approach, it seems
difficult to extract the physical content of the full quantum theory.
Finally, returning to canonical quantization, there are strong
indications that, in the final picture, all reference to a continuum
space-time would be lost, whence the notion of covariance may not be
meaningful at a fundamental level. The final picture may be more
combinatorial than geometric. The issue of covariance would be
meaningful only in the classical limit where it can be recovered
through the Hamiltonian description. The fact that this recovery is
not as direct as one would like is then largely an aesthetic question.
Furthermore, it is not unreasonable to hope that there may well exist
a better way of taking the classical limit --e.g. by exploiting the
equivalence${}^7$ between the traditional and covariant Hamiltonian
formulations of general relativity-- where covariance is manifest.

The approach I will discuss is being pursued by a large number of
researchers in about a dozen different research groups. I will not
attempt to present a comprehensive or even a systematic survey.
Rather, I will focus only on some of the recent mathematical
developments and show how these address certain concerns about the
viability of this specific program. Since there already exist a number
of excellent reviews on the subject, I will only sketch the main ideas
and provide references where further details can be found. Finally,
readers who are familiar with the program can skip section 2 and
proceed directly to sections 3 and 4 where the recent results are
discussed.

\vglue 0.6cm
\leftline{\twelvebf 2. Program}
\vglue 0.4cm

The program I want to discuss here has three conceptual ingredients:
i) A reformulation of Hamiltonian general relativity as a dynamical
theory of connections, which has close structural similarities with
gauge theories${}^{8}$; ii) A quantization of theories of connections
based on loops${}^{9,10}$; and, iii) An algebraic extension of Dirac's
treatment of constrained systems${}^{11,12}$. In this section, I will
briefly summarize the development of the program, roughly in the
chronological order. (For further details, see, e.g., Ref. [13-16].)

The canonical formulation of general relativity was first achieved in
the late fifties and early sixites through a series of papers by
Bergmann, Dirac and Arnowitt, Deser and Misner (ADM). In this
formulation, general relativity arises as a dynamical theory of
3-metrics. The framework was therefore named {\it geometrodynamics} by
Wheeler and used as a basis for canonical quantization both by him and
his associates and by Bergmann and his collaborators. The framework of
geometrodynamics has the advantage that classical relativists have a
great deal of geometrical intuition and physical insight into the
nature of the basic variables --3-metrics $q_{ab}$ and extrinsic
curvatures $K_{ab}$.  For these reasons, the framework has played a
dominant role, e.g., in numerical relativity. Unfortunately, it also
has two important drawbacks.  First, it sets the mathematical
treatment of general relativity quite far from that of theories of
other interactions where the basic dynamical variables are connections
rather than metrics. Second, the equations of the theory are rather
complicated in terms of metrics and extrinsic curvatures; being
non-polynomial, they are difficult to carry over to quantum theory
with any degree of mathematical precision. Consequently, as far as
full quantum general relativity is concerned, the work in this area
remained formal and concrete progress was restricted largely to
truncated models --minisuperspaces-- where all but a finite number of
degrees of freedom are frozen.

In the first step of the new approach, one performs a canonical
transformation on the phase space of geometrodynamics. The new
configuration variable is a (complex-valued) SU(2) connection,
$A_a^i$, the restriction to the 3-slice of the self-dual part of the
Lorentz, Spin connection of the 4-dimensional theory. The conjugate
momentum variable is a (density weighted) triad, $E^a_i$, on the
3-slice. These are related to geometrodynamical variables
non-polynomially. The 3-metric $q_{ab}$ is (apart from certain density
weights) the square of the triad $E^a_i$. The connection $A_a^i$ is
related to the spin connection $\Gamma_a^i$ of the triad via $A_a^i =
\Gamma_a^i - i K_a^i$, where $K_a^i$ is obtained by transforming
a spatial index on the extrinsic curvature $K_{ab}$ to an internal
index using the triad. This procedure casts general relativity as a
dynamical theory of connections $A_a^i$. Indeed, the phase space of the
theory is the same as that of Yang-Mills theory and the constraint
surface of general relativity is embedded into that of the Yang-Mills
theory.  Finally, the equations of the theory simplify significantly:
they are all low order polynomials in the new canonical variables.
Thus, the two drawbacks of geometrodynamics, mentioned above, have been
overcome. Furthermore, Einstein evolution is realized as a motion along
null geodesics in the infinite dimensional space of connections $A_a^i$.

There is, however, a price paid in this procedure: one has to impose
certain ``reality conditions'' on the canonical variables to ensure
that we are dealing with real, Lorentzian general relativity.  In
geometrodynamics, the reality conditions are trivial: the metrics and
the extrinsic curvatures are both required to be real. A self-dual
connection, on the other hand can be real only in the Euclidean
sector.  Therefore, if we demand that the new canonical variables be
real, we are led to Euclidean general relativity. If we let both
canonical variables be complex, we are led to complex general
relativity.  The reality conditions required to recover the Lorentzian
theory are rather subtle. To give an analogy from particle mechanics,
if we let $q$ and $p$ be the analogs of metrics and extrinsic
curvatures of geometrodynamics (which are both real), the new pair of
canonical variables, $(A_a^i, E^a_i$), is analogous to the pair $(z= q
-ip, q)$.

The second main ingredient in the program is the use of loop variables
for quantization of theories of connections%
\footnote{${}^\dagger$}{The loop representation was introduced by
Gambini and Trias${}^9$ for gauge theories and, independently, by
Rovelli and Smolin${}^{10}$ for general relativity. Recently, Gambini
and collaborators have introduced an extended loop representation.
(See [16] and references therein.)}.
The heuristic idea is as follows. In the Hamiltonian formulation,
quantum states of theories of connections arise as suitable
functionals $\Psi[A]$ of gauge equivalence classes of connections.
Now, consider the Wilson loop variables, $W_\alpha [A] := {\rm Tr} \
U_\alpha (A)$, where the group element $U_\alpha (A)$ is the holonomy
of the connection $A$ around the loop $\alpha$. Since $W_\alpha[A]$ is
the non-Abelian analog of $\exp i\oint_\alpha A.dl$, one can attempt
to define a generalized Fourier transform: $$ ``\ \ \psi(\alpha) :=
\int_{\ag} W_\alpha[A]\ \Psi[A]\ d\mu \ \ ''
\eqno(1)$$
where $d\mu$ is a measure on $\ag$ the space of connections modulo
gauge transformations. (The quotation marks denote that the expression
is formal since we have not specified the measure yet.) One can thus
take quantum states to be suitable functions of loops and define
operators directly on them. This turns out to be possible. Physically
interesting operators have simple actions that involve breaking,
re-routing and gluing various loops. This representation seems
especially well-suited for diffeomorphism invariant theories. In
general relativity, for example, one can impose the diffeomorphism
constraint on the loop states $\psi(\alpha)$ by asking that they
depend not on individual loops, but rather only on equivalence classes
of loops where two are equivalent if they are related by a
diffeomorphism${}^{10}$.  That is, it appears that the general
solution of the diffeomorphism constraint would be given by arbitrary
functions of (generalized) knot classes on the 3-manifold we began
with! This is an appealing idea and generated the initial enthusiasm
for loop representations in quantum gravity.  Furthermore, it turned
out that, on a similar heuristic level, a number of solutions to the
Hamiltonian constraints could be found. Some of them${}^{17}$ have
intriguing relations to well-known knot invariants that arise also in
the Chern-Simons theory.

The loop quantization methods have been extended to
supersymmetric${}^{18}$ and other matter couplings${}^{19}$.  The
``issue of time'' in quantum gravity was also analysed by coupling the
theory to a massless scalar field${}^{20}$. More precisely, in a
certain truncation, the classical theory was first deparametrized
using the scalar field as the time variable and the resulting model
was then quantized in the loop representation. This model admits a
true Hamiltonian and a number of interesting Dirac observables.  In
the spatially compact case, for example, it appears that the total
volume of space is quantized. Finally, certain approximation methods
have been developed to handle the low energy regime${}^{4, 21}$.  In
particular, gravitons arise as approximate notions${}^{22}$, the
approximation becoming better at long wave lengths.

This route to quantization has several unfamiliar features. First, the
basic canonical variables are hybrid; the connection is complex while
the triad is real. Second, the configuration variables of the theory
--the Wilson loop variables $W_\alpha$-- are {\it over}complete; there
are identities between them. In the Dirac quantization scheme, there
is no obvious prescription to handle these identities in the quantum
theory.  The overcompleteness also holds for momentum variables $P_S$
which are associated with 2-dimensional (ribbons or) strips $S$.
(Recall that the Wilson loop variables $W_\alpha$ are associated with
loops $\alpha$.)  Third, there are the awkward reality conditions
which should be incorporated in quantum theory. Finally, in
Minkowskian quantum field theories, one uses the Poincar\'e group to
select the vacuum state and constrain the inner product. In the
present case, we do not have a background structure and hence there is
no obvious symmetry group. (Note that diffeomorphisms act as gauge and
therefore have a trivial action on physical states.) Thus, we need a
new principle to constrain the Hilbert space structure.

To handle such issues, an algebraic extension of Dirac's method of
quantization was carried out and applied to a variety of simpler
models which share one or more of the above features with the
connection formulation of general relativity. These models include, in
particular, some minisuperspaces of 4-dimensional general relativity,
linearized gravity and 3-dimensional general relativity. In all these
cases, the unusual features of connection dynamics could be addressed
and the extended program could be completed. Thus, the algebraic
extension of the Dirac program provides a consistent framework to
incorporate loop quantization of the connection dynamics formulation
of general relativity.

Finally, the general framework has had certain applications to high
energy physics${}^{23,24}$ and to problems in geometry${}^{25}$.

\vglue 0.6cm
\leftline{\twelvebf 3. Concerns}
\vglue 0.4cm

While the program I just outlined is structurally coherent, there are,
nonetheless, a number of issues of mathematical precision that were
left open initially. This is of course unavoidable in any initiative
--such as loop quantization-- which opens up previously unexplored
directions. However, a number of years have passed since the initial
burst of activity and it is now appropriate to attempt to weed out the
technically unsound results and make the foundations firm so as to
build on them further.

Over the past couple of years, several concerns have been expressed
about some of the main results of the program. First, there is the
issue of the meaning of the loop transform of Eq (1). What is the
meaning of the measure used in this transform? If the measure is only
formal, will not the operators on loop states be ambiguous?  One can
take the standpoint that the loop representation is the fundamental
one and the transform is only a heuristic devise. That is, one might
just begin with a vector space $\V$ of ``kinematical'' loop states
$\psi(\alpha)$ and define on it directly operators $\hat{W}_\alpha$
and $\hat{P}_S$ corresponding to the classical configuration and
momentum observables, ensuring that ($i\hbar$ times) the classical
Poisson brackets go over to commutators. While this strategy is
attractive --and was in fact followed-- it does not eliminate the
problems.

First, classically, the loop-strip observables are subject to a
variety of relations (some of which are inequalities). One cannot
simply ignore these if, e.g., one wants the theory to have the correct
classical limit. (This problem emerges rather clearly in the loop
representation of 3-dimensional quantum general relativity.${}^{26}$.)
If the loop representation is constructed using a transform, these
relations are automatically incorporated. In the more direct approach
to loop states, the task of incorporating all these relations seems
quite difficult.

Then there is the question of constraint operators. Since we are
dealing with a field theory, these have to be regulated appropriately.
Without a good deal of control on the vector space $\V$, this is hard
to achieve.  Therefore, in practice, one proceeded as follows. One
simply imposed those regularity conditions on loop states that seemed
to be essential for the regularization procedure to be meaningful.
Overall, this is a reasonable strategy. One {\it can} criticize it on
the grounds that the rules of the game are not clear apriori in the
sense that there is no underlying conceptual framework that would tell
us which regularity conditions are reasonable and which are not.
However, this criticism can be made also against many of the
regularization procedures used in Minkowskian field theories. Indeed,
the overall level of mathematical precision in some of the heuristic
treatments of loop quantization is comparable to that used routinely
in field theory. However, there is an important difference between the
two situations. We have a great deal of theoretical experience with
the procedures used in Minkowskian field theories and, more
importantly, we have a lot of experimental justification backing them.
In non-perturbative quantum gravity, we have neither! We are in
completely new territory and it is therefore all the more important
that we reduce the number of ad-hoc steps.  For example, in field
theory, although we may begin with smooth configurations, quantum
states turn out to be functionals of distributional fields.  Indeed,
the smooth configurations constitute a set of measure zero!  So, are
we justified even in the assumption that the loop states should be
functionals of ``nice'' (say piecewise analytic) loops?  Shouldn't we
consider more general loops? If we do, it would be very hard to impose
the required regularity conditions on $\V$. Furthermore, some of the
regularity conditions require that values $\psi(\alpha)$ of wave
functions on individual loops be well-defined. Now, already in
non-relativistic quantum mechanics, whether wave functions in the
physical Hilbert space take on well-defined values at {\it every}
point of its argument depends on the choice of representation. In the
Schr\"odinger representation, they need not, while in a holomorphic
representation they do. What is the situation for the loop
representation?

These concerns in turn raise questions about the solutions to quantum
constraints, i.e., about the space $V_{\rm phy}$ of physical states of
quantum gravity. How reliable are the results on solutions to
constraints? In solving constraints, no (uncontrolled) infinities were
encountered so far. Is this simply because we are ignoring them?
Afterall, if we restrict ourselves to smooth configurations for, say
QED, there would be no infinities in the potential term of the quantum
Hamiltonian. Is something similar happening here? Are we throwing away
anomalies by hand? Is the result on the general solution to the
diffeomorphism constraint accurate? Can we really have loop states
which are characteristic functions of knots?

Finally, there is the important question of reality conditions. In the
heuristic treatments, the following strategy is adopted. One simply
ignores the reality conditions initially, thereby considering, in
effect, complex general relativity. The idea is to impose reality
conditions on physical observables at the end, thereby constraining
severely the form of the inner product on physical states. Again, the
strategy itself is conceptually sound and enabled one to make
considerable progress. However, now that these methods have yielded a
number of solutions to quantum constraints, before going too far along
this road, it is appropriate to ask if there are indications that
these solutions refer to the real, Lorentzian sector. And, there {\it
are} reasons to be concerned that the simplest solutions obtained so
far may not refer to this sector. For example, in the classical
theory, using connection dynamics, it is easy to write down an
infinite family of solutions to all constraints${}^{27}$. But these
correspond to self-dual solutions and the only ones that satisfy the
Lorentzian reality conditions correspond to initial data for Minkowski
space! Is something similar happening in the quantum theory as well?
Is there a way to analyse this issue now, without having to first
solve the full theory? One strategy would be to incorporate the
reality conditions at the kinematic level, i.e., translate to quantum
theory the property that the constraint functions are real in real
general relativity.  Is there a rigorous procedure to implement this
strategy?

\vglue 0.6cm
\leftline{\twelvebf 4. Recent Developments}
\vglue 0.4cm

In this section, I will summarize some recent mathematical
developments that provide a sharper focus to the program thereby
putting it on a coherent and sound footing. In particular, we will see
that most of the concerns raised above have now been addressed.

1. In a systematic treatment, one can begin with geometrodynamics and
push it as far as possible. For this, one can use the real-valued,
$SU(2)$ spin connection $\Gamma_a^i$ --constructed solely from the
triad $E^a_i$-- as the configuration variable. This is only a
reformulation of geometrodynamics but has the advantage that one can
now employ all the mathematical machinery available to deal with
theories of connections.  This strategy resolves some of the important
issues in {\it quantum} geometrodynamics. As we will see, one can go
quite far along these lines. However, to deal with the Hamiltonian
constraint, it will be necessary to go to the (self-dual) complex
connections $A_a^i$. (see point 5 below.)

2. Denote by ${\cal A}$ the space of all $SU(2)$ connections
$\Gamma_a^i$ which are compatible with a non-degenerate triad $E^a_i$.
The space $\agr$ of gauge equivalence classes of such connections is
now the classical configuration space of the theory.  (It is
essentially the same as the space of positive definite metrics.)  A
key question for {\it quantum} geometrodynamics is: What is the
quantum configuration space? More precisely, in the connection
representation, what is the domain space of quantum wave functions?
In quantum mechanics --i.e., in the quantum theory of systems with a
finite number of degrees of freedom-- the classical and the quantum
configuration spaces agree. In field theory, by contrast, the quantum
configuration space is a substantial enlargement of the classical one;
in scalar field theories, for example, although the classical
configurations are smooth (say $C^2$) functions on a $t={\rm const}$
slice, the quantum configuration space consists of all tempered
distributions.  It turns out that there is a systematic way to answer
this question. The answer for geometrodynamics is the
following${}^{28-32}$: the quantum configuration space is a compact,
Hausdorff space $\agb$. It is a completion of $\agr$, obtained by
looking at the kinematics of the continuum theory as a precise
(namely, projective) limit of that of lattice gauge theories. For a
lattice with $n$ independent loops, the (classical as well as the
quantum) configuration space is $[SU(2)]^n$.  $\agb$ is a projective
limit of $[SU(2)]^n$. As expected, the extended configuration space
does admit generalized connections which are ``distributional''.

3. To proceed with quantum theory, one has to specify an inner product
on the space of suitable functionals on $\agb$. This can be achieved
by specifying measures on $\agb$. Note that $\agb$ does admit regular
measures in a rigorous sense; it a compact, Hausdorff space. One can
show that regular measures $\mu$ on $\agb$ are in 1-1 correspondence
with certain functions $\chi_\mu$ on the space of (based) loops on the
given 3-mainfold. To see this, note first that the construction of
$\agb$ is such that the Wilson loop functions $W_\alpha[A]$ admit
natural extensions $W_\alpha[\Ab]$ to $\agb$. These functionals are
continuous and bounded on $\agb$. Hence, given any regular measure
$\mu$ on $\agb$, we can integrate them. The result is the required
function $\chi_\mu$ on the loop space: $$ \chi_\mu (\alpha) =
\int_{\agb} W_\alpha [\bar{A}]\ d\mu[\bar{A}]
\eqno(2)$$
$\chi_\mu$ is called the generating functional for the measure $\mu$.
Conversely, a function $\chi$ on the loop space  defines a regular
measure with $\chi$ as its generating function provided it satisfies
the following two conditions:
\item{i)}{ $\sum_{i=1}^{n} k_i\ \chi(\alpha_i) = 0$ if
 $\sum_{i=1}^{n} k_i\ W_{\alpha_i} [A] = 0, \forall [A]$; and}
\item{ii)} {$\sum_{i,j=1}^{n} \bar{k}_i k_j\ (\chi(\alpha_i\circ\alpha_j)
+ \chi(\alpha_i^{-1}\circ\alpha_j) \ge 0$}

\noindent for all complex numbers $k_i$ and integers $n$.
(In the second condition, the loops are composed at the base point.)
The first condition ensures that we have handled the overcompleteness
of the Wilson loop observables satisfactorily in the quantum theory,
while the second ensures that the expectation value of a positive
operator is positive. Eq.(2) is the rigorous analog of the heuristic
loop transfrom of Eq.(1). Thus, $\chi_\mu(\alpha)$ are the quantum
states in the loop representation. Note that while states in the
connection representation have support on generalized
(``distributional'') connections, the loop states are functions of
{\it ordinary}, nice (more precisely, piecewise analytic) loops. Thus,
there is no need to consider ``distributional loops''. Finally, by
construction, $\chi_\mu$ takes on a well-defined value on every loop
$\alpha$, whence, it {\it is} meaningful to formulate regularity
conditions in terms these values.

We can construct measures on $\agb$ explicitly. There is one measure
$\mu_o$ in particular which is natural in the sense that (in the
projective limit) it is induced simply by the Haar measure on
$SU(2)$. It does not require any additional input, is strictly
positive and diffeomorphism invariant.  It is natural to use it to
construct a kinematical Hilbert space $H_{\rm kin} = L^2(\agb ,
d\mu_o)$, the precise analog of $\V$ of section 3. On this Hilbert
space, one can define the configuration and momentum operators
$\hat{W}_\alpha$ and $\hat{P}_S$ and show that they are self-adjoint.
Since classically, $W_\alpha$ and $P_S$ are real and constitute a
complete set of observables, the self-adjointness implies that we have
incorporated the classical, kinematical ``reality conditions'' in
quantum kinematics.

4. The next task is to formulate and solve the diffeomorphism
constraint. It turns out that the momentum $P^a_i$, conjugate to
$\Gamma_a^i$ is related to the triad and the extrinsic curvature in a
non-local fashion. Nonetheless, one can express each diffeomorphism
constraint as an operator on $\H$. The commutator algebra of these
operators is the same as that in the classical theory. There are no
anomalies. As one might expect, the solutions of the diffeomorphism
constraint are not elements of $\H$. This is a common occurrence even
in particle mechanics. There is, however, a precise treatment of the
problem and the natural home for solutions is the space of regular,
complex-valued (i.e., ``signed'') mesures $\mu$ on $\agb$.
Alternatively, the home for solutions is the space of functions
$\chi_\mu$ determined via Eq.(2) by a complex-valued measure $\mu$.
In this space, we can seek solutions. Not surprisingly, they are
simply diffeomorphism invariant measures on $\agb$. Alternatively,
they are functions $\chi(\alpha) $ on the loop space which depend only
on the generalized%
\footnote{${}^\dagger$}{\ninerm \baselineskip=7pt\noindent
We will call a knot {\it regular} if its representative loops are
smoothly embedded in the 3-manifold and {\it generalized} if they are
not (e.g., if they have intersections, kinks or overlaps).}
knot class of $\alpha$ {\it and satisfy the two algebraic conditions}
which qualify them as generating functions of signed measures.

These conditions are important; contrary to what one might expect at
first, {\it not every knot-invariant can solve the diffeomorphism
constraint!} In particular, we cannot set $\chi(\alpha)$ to be a
characteristic function of a regular knot class. One {\it can},
however, construct solutions which satisfy all these requirements. In
particular, given a suitable invariant $k(\alpha)$ of regular knots,
using the fiducial measure $\mu_o$ one can generate a new
(complex-valued) measure $\mu_k$ on $\agb$ which is also
diffeomorphism invariant. The generating function $\chi_k(\alpha) $ of
this measure $\mu_k$ is then a solution to the diffeomorphism
constraints. However, it differs from $k(\alpha)$; in particular, it
does not necessarily vanish on loops with intersections, kinks or
overlaps. Thus, while the ``obvious'' results on the general solution
to the diffeomorphism constraint is actually incorrect, its general
spirit {\it is} realized in the rigorous result.

5. Thus, within {\it quantum} geometrodynamics, we have
answered${}^{33}$ three questions: i) What is the domain space of
quantum states? (answer: $\agb$); ii) Are there anomalies in a
rigorous treatment of the diffeomorphism constraint?  (answer: no);
and, iii) Can one write down the ``general solution'' rigorously?
(answer: yes). The first and the third questions were raised by
Wheeler already in the seventies. (There was an implicit assumption
that the second question would be answered affirmatively.) However,
the subsequent analysis was carried out within the Hamiltonian
formulation of the {\it classical} theory. Parallel developments in
quantum field theory taught us that, in systems with an infinite
number of degrees of freedom, the structure of the quantum
configuration space is quite different from that of the classical
configuration space and that this difference has to be faced squarely.
(This point has been emphasized by Isham ${}^{34}$.).  The present
treatment takes this lesson seriously. The upshot is that the {\it
quantum} analog of Wheeler's superspace is the space of diffeomorphism
invariant, signed measures $\mu$ on $\agb$, or equivalently, their
generating functionals $\chi_\mu$ on the loop space.

 From a structural point of view, our main remaining task is to treat
the Hamiltonian constraint in a similar fashion. As one might expect,
this constraint does not seem manageable within geometrodynamics in
the sense that there are no realistic ideas on how one might take it
over to the quantum level. {\it It is here that we need to go to the
complex (self-dual) connections} $A_a^i$: we know that the classical
Hamiltonian constraint does simplify considerably in terms of $(A_a^i,
E^a_i)$.  However, if one works directly in these variables, one has
to face the issue of reality conditions all over again. Fortunately,
this problem can be avoided by constructing a transform from the real
connection ($\Gamma_a^i$-) representation to the complex connection
($A_a^i$-) representation.

The idea is to mimick the celebrated Segal-Bargmann transform${}^{35}$
which maps one from the Schrodinger representation where the wave
functions are arbitrary, complex-valued functions of a real variable
$q$ to the coherent state representation where they are {\it
holomorphic} functions of a {\it complex} variable $z= q-ip$. This
transform can be constructed using heat kernel methods.  Using some
key recent results due to Hall${}^{36}$, this transform has been
extended${}^{37}$ to the case of general relativity under
consideration. The result is a rigorous, quantum field theoretic
analog of the canonical transformation that took us from classical
geometrodynamics to (self-dual) connection dynamics.  Since the two
representations are isomorphic, the kinematical reality conditions
which were incorporated in the real connection representation, are
incorporated also in the holomorphic representation. Thus, what we
have is a kinematic arena for dealing with all constraints, where the
Wilson loop functions of the complex connection $A_a^i$ act by
multiplication.

The discussion of the diffeomorphism constraint extends in a
straightforward way to the holomorphic representation and the final
results are completely analogous to the one discussed above.  Work on
the Hamiltonian constraint has only begun. There is a rich body of
heuristic results on the treatment of this constraint (see, e.g., Ref.
[10, 13-20, 38]). The hope is that they would lead to a precise
expression of the constraint in the holomorphic Hilbert space. If this
turns out to be the case, one would say that quantum general
relativity does exist non-perturbatively. One can then apply suitable
approximation techniques to extract the physical content of the
theory. Such methods are already being developed at a heuristic
level.${}^{18-21}$ Indeed, the strength of the program lies in this
simultaneous development of the heuristic techniques and the rigorous
framework, each stimulating further developments in the other.

\vglue 0.6cm
\leftline{\twelvebf 5. Discussion}
\vglue 0.4cm

Let us briefly summarize the overall status.

There now exists a rigorous kinematical framework to deal with quantum
gravity non-perturbatively. In terms of physical predictions, this
progress is modest.  However, these developments have provided us
quite a different language to analyze various issues. And the answers
have a new degree of precision. Furthermore, even the formulation of
some of the basic questions has altered. An outstanding example is:
How do you impose constraints?  While the quantum constraints
$\hat{C}$ arise as operators on the kinematical Hilbert space $\H$,
one imposes them on {\it measures} $\mu$ on $\agb$ by demanding: $$
\int_{\agb} (\hat{C}\circ W_\alpha[\Ab]) \ d\mu = 0\ , \forall
\alpha \  . \eqno(3)$$
The issue of regularization one faces in this formulation is rather
different from that in the standard Dirac picture. Similarly, new
possibilities for solutions arise. For example, one would expect that,
in the holomorphic representation, measures which have support on the
moduli space of flat connections would automatically satisfy all
constraints. These solutions would carry information about the
topology (the first homotopy group) of the manifold.  It is difficult
to transcribe this idea in the Dirac language of imposing constraints
on {\it functions} on $\agb$. More generally, the language provided by
the recent developments enables new constructions and provides new
translations of physical questions as well as methods for tackling
them.  Indeed, the situation has a certain degree of similarity to the
emergence of global techniques in classical general relativity which
also provided a new language and new ways of thinking of physical
issues such as singularities and black holes.  The local, coordinate
dependent techniques were certainly useful and they did lead to many
interesting results. However, their power was restricted.  Modern
differential geometry added a new degree of precision which in turn
resolved certain confusion and led to a variety of powerful results.
One would hope that the transition from heuristic methods to the ones
discussed in the last section would bring similar fruits.

Although the developments which led to the new framework are
mathematical, the various constructions involved do have physical
counterparts. The use of loops to probe function spaces of
connections, for example, leads one to a theory in which the
fundamental excitations of the gravitational field are ``loopy''
rather than wave-like. These in turn lead to a discrete picture of
quantum geometry and a theory in which the fundamental operators act
via combinatorics rather than functional derivatives.  The basic
interactions will not be mediated by gravitons; they will correspond
to breaking, re-routing and gluing of loops. Thus, the framework is
{\it fundamentally} different from the perturbative one.

\vglue 0.6cm
\leftline{\twelvebf 5. Acknowledgements}
\vglue 0.4cm
The new results reported in this article were obtained in
collaboration with Jerzy Lewandowski, Donald Marolf, Jos\'e Mour\~ao
and Thomas Thiemann. I would like to thank them as well as Jorge
Pullin, Carlo Rovelli and Lee Smolin for innumerable discussions. This
research was supported in part by the NSF grant PHY93-96246 and by the
Eberly research fund of Penn State University.

\vglue 0.6cm
\leftline{\twelvebf 6. References}
\vglue 0.4cm

\itemitem{1.} P. Faria de Veiga, Ecole Polytechnique thesis (1990);
C. de Calan, P. Faria de Veiga, J. Magnen and R. S\'en\'eor, {\tit
Phy. Rev. Lett.} {\tbf 66} (1991) 3233; A. S. Wightman, in: {\twelveit
Mathematical Physics Towards XXIst Centrury}, edited by R. N. Sen and
A. Gersten (Ben Gurion University Press, 1994).
\itemitem{2.} D. Amati, M. Ciafolini and G. Veneziano, {\tit Nucl. Phys.}
{\tbf B347} (1990) 550.
\itemitem{3.} A. Ashtekar, C. Rovelli and L. Smolin, {\tit Phys. Rev.
Lett.} {\tbf 69} (1992) 237.
\itemitem{4.} A. Agishtein and A. Migdal, {\twelveit Mod. Phys. Lett.}
{\tbf 7} (1992) 85.
\itemitem{5.} A. Ashtekar and M. Varadrajan, {\tit Phys. Rev.} {\tbf D}
(in press).
\itemitem{6.} A. Ashtekar, J. Lewandowski, D. Marolf, J. Mour\~ao and
T. Thiemann  (in preparation).
\itemitem{7.} A. Ashtekar and A. Magnon, {\tit Commun. Math. Phys.}
{\tbf 86} (1982) 55.
\itemitem{8.} A. Ashtekar, {\tit Phys. Rev. Lett.} {\tbf 57} (1986) 2244;
{\tit Phys. Rev.}{\tbf D36} (1987) 1587.
\itemitem{9.} R. Gambini and A. Trias, {\tit Phys. Rev.} {\tbf D22} (1980)
1380; {\tit Nucl. Phys.} {\tbf B278} (1986) 436.
\itemitem{10.} R. Rovelli and L. Smolin {\tit Nucl. Phys.} {\tbf B331}
(1990) 80.
\itemitem{11.}A. Ashtekar, {\tit Non-Perturbative Canonical Gravity}
(World Scientific, Singapore, 1991).
\itemitem{12.} A. Ashtekar and R. S. Tate, {\tit J. Math. Phys} (in press).
\itemitem{13.} R. Rovelli, {\tit Class. and Quantum Grav.} {\tbf 8} (1991)
1613.
\itemitem{14.} L. Smolin, in {\tit The Proceedings of the XXII Gift
International Seminar} (World Scientific, Singapore, 1992).
\itemitem{15.} A. Ashtekar, in {\it Gravitation and Quantizations} ed.
B. Julia (Elsevier, Amsterdam 1994).
\itemitem{16.} R. Gambini, in {\tit Proceedings of the IVth Mexican
Workshop on Particles and Fields} (World Scientific, Singapore, in
press).
\itemitem{17.} B. Br\"ugmann, R. Gambini and J. Pullin, {\tit Phys. Rev.
Lett.} {\tbf 68} (1992) 431.
\itemitem{18.} H. Nicolai and H. J. Matschull, {\tit J. Geo. and Phys.}
{\tbf 11} (1993) 15; H. J. Matschull, pre-print gr/qc 9305025;
\itemitem{19.} H. A. Morales-T\'ecotl and C. Rovelli, {\tit Phys. Rev.
Lett.}{\tbf 72} (1994) 3642.
\itemitem{20.} C. Rovelli and L. Smolin, {\tit Phys.  Rev. Lett.}
{\tbf 72} (1994) 446.
\itemitem{21.} J. Zegwaard {\tit Phys. Lett.} {\tbf B378} (1993) 217.
\itemitem{22.} J. Iwasaki and C. Rovelli, {\tit Int. J. Mod. Phys.}
{\tbf D1} (1993) 533; {\tit Class. \& Quantum Grav.}{\tbf 11} (1994)
1653.
\itemitem{23.} A. Ashtekar, A. P. Balachandran and S. G. Jo, {\tit Int.
J. Mod. Phys.} {\tbf  A4} (1989) 1493; L. Chang and C. Soo, pre-print
CGPG-94/6-2.
\itemitem{24.} B. Br\"ugmann, {\tit Phys. Rev.} {\tbf D43} (1991) 566;
A. Ashtekar, J. Lewandowski, D. Marolf and T. Thiemann, in {\tit Geometry
of Constrained Dynamical Systems} ed. J. Charap (Cambridge University
Press, Cambridge, 1994).
\itemitem{25.} J. Samuel, in {\tit Recent Advances in General Relativity},
ed. A. Janis and J. Porter (Birkh\"auser, Boston, 1991).
\itemitem{26.} D. Marolf, {\tit Class. and Quantum Grav.} {\tbf 10} (1993)
2625; A. Ashtekar and R. Loll, {\tit Class. and Quantum Grav.} (in press).
\itemitem{27.} A. Ashtekar, T. Jacobson and L. Smolin {\tit Commun. Math.
Phys.} {\tbf 11} (1988) 631.
\itemitem{28.} A. Ashtekar and C. J. Isham, {\tit Class. and Quantum Grav.}
{\tbf 9} (1992) 1433.
\itemitem{29.} A. Ashtekar and J. Lewandowski, in {\tit Knots and Quantum
Gravity}, ed. J. Baez (Oxford University Press, Oxford, 1994).
\itemitem{30.} J. Baez, {\tit Lett. Math. Phys.}{\tbf 31} (1994) 213;
in {\tit The Proceedings of the Conference on Quantum Topology},
ed D. N. Yetter (World Scientific, Singapore, in press).
\itemitem{31.} D. Marolf and J. Mor\~ao, preprint, CGPG 94/3-1.
\itemitem{32.} A. Ashtekar, D. Marolf and J. Mor\~ao, in {\tit
The Proceedings of the Lanczos International Centenary Conference},
ed. D. Brown et al (SIAM publishers, Philadelphia, 1994).
\itemitem{33.} A. Ashtekar, J. Lewandowski, D. Marolf, J. Mour\~ao and
T. Thiemann, {\tit Quantum geometrodynamics} (preprint)
\itemitem{34.} C. J. Isham, in {\tit Relativity, Groups and Topology, II},
ed B. S. DeWitt and R. Stora (North Holland, Amsterdam, 1984).
\itemitem{35.} I.E. Segal, in {\tit Proceedings of the Summer Conference,
Boulder, Colorado} ed M. Kac (1960); V. Bargmann, {\tit Commun. Pure and
App. Math.} {\tbf 24} (1961) 187.
\itemitem{36.} B.C. Hall, {\tit J. Funct. Analysis} (in press).
\itemitem{37.} A. Ashtekar, J. Lewandowski, D. Marolf, J. Mour\~ao and
T. Thiemann, {\tit Coherent state transfrom on the space of connections}
(preprint)
\itemitem{38.} B. Br\"ugmann and J. Pullin, {\tit Nucl. Phys.} {\tbf B363}
(1991) 221; {\tbf B390} (1993) 399.

\end\bye

J. F. Donoghue and B. R. Holstein, {\twelveit Phys.  Rev.} {\twelvebf
D25} (1982) 2015.
\itemitem{2.} M. L. Cohen and P. W. Anderson, in {\twelveit
Superconductivity in d- and f-Band Metals}, ed. D. H. Douglas
(AIP, New York, 1972).
\itemitem{3.} H. Krebs, {\twelveit Fundamentals of Inorganic Crystal
Chemistry} (McGraw-Hill, London, 1968), p. 160.

\vglue 0.6cm
\leftline{\twelvebf 7. Footnote}
\vglue 0.4cm
Footnotes should be typeset in 9 point roman at the bottom of
the page where it is cited.
\bye